# Optimization of broadband omnidirectional antireflection coatings for solar cells

Xia Guo[1], Qiaoli Liu[1], Chong Li[1]*, Hongyi Zhou[1], Benshun Lv[1], Yajie Feng[1], Huaqiang Wang[1] and Wuming Liu[2]

[1]Institute of Electronic Information and Control Engineering, Beijing University of Technology, Beijing 100124, China,

[2]Beijing National Laboratory for Condensed Matter Physics, Institute of Physics, Chinese Academy of Sciences, Beijing 100190, China

Broadband and omnidirectional antireflection coating is a generally effective way to improve solar cell efficiency, because the destructive interference between the reflected and input waves could maximize transmission light in the absorption layer. Several theoretical calculations have been developed to optimize the anti-reflective coating to maximize the average transmittance. However, the solar irradiances of the clear sky spectral direct beam on a receiver plane at different positions and times are variable greatly. Here we report a new theoretical calculation of anti-reflective coating with incident quantum efficiency $\eta_{in}$ as evaluation function for practical application. The two-layer and three-layer anti-reflective coatings are optimized over $\lambda$ = [300, 1100] nm and $\theta$ = [0°, 90°] for cities of Quito, Beijing and Moscow. The $\eta_{in}$ of two-layer anti-reflective coating increases by 0.26%, 1.37% and 4.24% for these 3 cities, respectively, compared with that other theoretical calculations due to better match between the local actual solar spectrum and quantum efficiency spectrum. Our numerical simulation and comparison data with other optimization methods suggest that this optimization method combining ant colony algorithm method with SPCTRL2 solar spectral irradiance can effectively push the efficient solar cell toward higher quantum efficiency, thus enabling high utilization efficiency of solar irradiance.

## Introduction

Broadband and omnidirectional antireflection is one of the most desirable performances for the high-efficiency solar cells to allow the maximum light transmission into the absorption layer, which can be obtained by a few additional layers on the outside of the Si solar cells that occupies about 80% in the photovoltaic market due to the mature process technology and low price[1]. As early as 1880, it was found that graded refractive-index multilayers increasing from top to bottom of the film between air and Si substrate have the broadband transmittance property both at normal and oblique incident angles[2,3,4,5]. The insect compound eyes are one of the typical graded refractive-index structures, which inspired lots of nanostructures and fabrication techniques developed over the last years, such as arraying of $SiO_2$ microspheres[6], Si nanowires or nanoparticles which are inspired by laser direct writing or nan-imprinting[7-8], nano-rods inspired by hetero-epitaxial growth[9,10,11,12,13,14], and etc.. In order to obtain the materials with a refractive index lower than 1.4 which are highly desirable for anti-reflective coatings, oblique-angle deposition technique was developed with the minimum refractive index being 1.05[15,16,17,18]. Recently, the solution of polymethyl methacrylate (PMMA) was spin-coated on a modified glass substrate to achieve a graded distribution of PMMA domains in the vertical direction of the entire micro-phase separated film[19]. The transmittance is above 97% both in the wavelength ranging from 400nm to 800nm and 800nm to 2000 nm [20,21,22,23].

In the view of the effective medium theory (EMT), the anti-reflective(AR) coatings based on the insect compound eyes formed by the nanoscale arrays with smoothly graded refractive index changing profile from air to substrate[24,25,26,27], light is tend to be bent progressively. And the polarization of the light is insensitive to such anti-reflective coatings with low disparity of refractive index, which is one of the important requirements for broadband and omnidirectional anti-reflective coatings [28]. The performance of the anti-reflective coatings with nanostructures depends on the etching process and functions as the function of volume fraction of inclusion for a material mixture [29,30,]. Generally, different refractive index profile curves, such as linear, parabolic, cubic, quantic, exponential and exponential-sinusoidal, exhibit different anti-reflective performance[31,32,33]. The optimization of the refractive-index distribution in the vertical direction is then changed to the optimization of the profile of nanostructure and its filling factor [34,35,36]. However, it is still a great challenge to obtain an optical structure with high performance, though it is easier to characterize its optical performance [37,38,39].Therefore, theoretical calculation becomes extremely important for optimizing high-performance anti-reflective coatings. Several rigorous optimization algorithms, such as genetic algorithm [40,41], simulated annealing algorithm [42, 43] and ant colony algorithm (ACA)[44], have been developed to optimize the anti-reflective coating system which can present the thickness and corresponding refractive-index distribution of the anti-reflective coatings. In previous results, the average reflectance of 2.98% optimized by ant colony algorithm [14] is obtained, which is much lower than 4.5% optimized by genetic algorithm [10] and 6.59% by simulated annealing algorithm [13] over the same optimization range of $\lambda$ = [400, 1100] nm and $\theta$ =[0°, 80°]. Moreover, From the viewpoint of the number of particles, instead of traditional energy during the characterization of reflectance or transmittance, the objective of optimizing the anti-reflective coatings is to obtain high incident quantum efficiency $\eta_{in}$ which is proportional to the external quantum efficiency of the solar cells, and defined as the ratio of the number of the photons passing through the anti-reflective coating into the solar cell to the total incident photon number shining on the solar cell from outside. Due to the mismatch between the uniform irradiation which was widely used in the above algorithm optimizations and actual solar spectrum which is wavelength and incident angle dependent, the actual quantum efficiency of solar cells must be lower than the calculated efficiency[13-14].

In this paper, the incident quantum efficiency $\eta_{in}$ is proposed to be the evaluation function with solar spectrum incorporated for the optimization of broadband omnidirectional anti-reflective coating by using ant colony algorithm method. SPCTRL2 [45], a program computing solar irradiance of the clear sky spectral direct beam on a receiver plane at a position and time, is involved during the optimization. Three typical cities of Quito, Beijing and Moscow, which locate at Equator, northern latitudes of 39.9° and 55.3°, respectively, are selected for anti-reflective coating optimization over $\theta$ =[0°, 90°]. And $\lambda$= [300, 1100] nm. The $\eta_{in}$ of two-layer anti-reflective coating optimized by ant colony algorithm with SPCTRL2 incorporated increase from 95.74% to 96.00% for Quito, from 92.27% to 93.64% for Beijing, and from 86.78% to 91.02% for Moscow. With three-layer anti-reflective coating, the $\eta_{in}$ increase from 96.86% to 97.60% at Quito, from 95.72% to 97.03% at Beijing, and from 94.46% to 95.74% at Moscow. The optimization of anti-reflective coating with $\eta_{in}$ as evaluation function can effectively improve the quantum efficiency of the solar cells.

**Results**:

**The incident quantum efficiency and photons flux density:** The optimization structure of solar cells with graded refractive-index anti-reflective coating is shown in Fig. 1, where a plane wave is incident from the air with an incident angle of $\theta$ and the refractive-index of $n_0$. Each layer is characterized by its thickness $d_i$ with the refractive-index $n_i$, i = {1, 2, …, N}. For simplicity, the entire absorption layer is assumed to be in the bottom absorption layer with the refractive-index of $n_{ab}$ without considering the back surface reflectance [46].

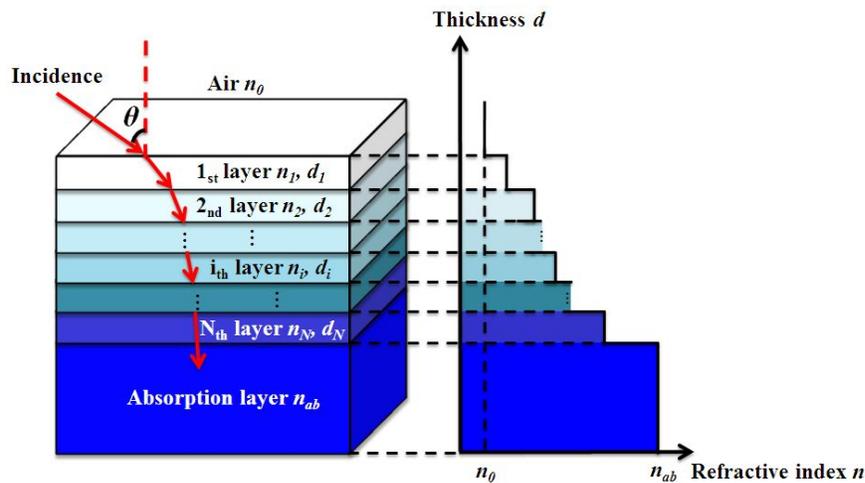

Fig. 1.The graded refractive-index structure of N-layer anti-reflective coating system. A plane wave is incident from the air with the refractive-index of $n_0$ and an incident angle of $\theta$. Every layer of the coating is characterized by the thickness $d_i$ with the refractive-index $n_i$, i = {1, 2, …, N}. The entire absorption layer is assumed to be in the bottom layer with the refractive-index of $n_{ab}$ without the back surface reflectance. The right curve shows numerical value and the graded change of refractive-index and the thickness of each anti-reflective layer.

One parameter of the incident quantum efficiency is broadband and omnidirectional incident quantum efficiency $\eta_{in}$, which is independent on the wavelength and angles, expressed by:

$$\eta_{in} = \frac{\int_{\lambda_1}^{\lambda_2} \int_{\theta_1}^{\theta_2} T(\lambda,\theta) Num(\lambda,\theta) \mathrm{d}\theta \mathrm{d}\lambda}{\int_{\lambda_1}^{\lambda_2} \int_{\theta_1}^{\theta_2} Num(\lambda,\theta) \mathrm{d}\theta \mathrm{d}\lambda} , \quad (1)$$

Another parameter is omnidirectional incident quantum efficiency $\eta_{in}(\lambda)$, which is dependent on the wavelength and can be calculated by the equation shown below:

$$\eta_{in}(\lambda) = \frac{\int_{\theta_1}^{\theta_2} T(\lambda,\theta) Num(\lambda,\theta) \mathrm{d}\theta}{\int_{\theta_1}^{\theta_2} Num(\lambda,\theta) \mathrm{d}\theta}, \quad (2)$$

in which $\lambda_2$ and $\lambda_1$ are the upper and lower boundaries of wavelength, $\theta_2$ and $\theta_1$ are the upper and lower boundaries of incident angle. $T(\lambda, \theta)$ is the optical transmittance of N-layers anti-reflective coating, which is calculated by the transmission matrix method according to the Maxwell equation [13, 14]. The $Num(\lambda, \theta)$ (photon/m2/s) is a function of incident wavelength $\lambda$ and the incident angle $\theta$, which can be expressed by [47, 48]:

$$Num(\lambda,\theta) = E(\lambda,\theta) \frac{\lambda}{hc}, \quad (3)$$

Wherein $h$ is the Planck's constant and $c$ is the speed of light in vacuum. $E(\lambda,\theta)$ (in units of W/m²/micron interval) is the irradiance of solar spectrum.

It is necessary to consider the local solar spectrum for a whole year where the solar cell systems will be located during the optimization of the anti-reflective coatings. Solar spectrum of AM1.5 is widely used to characterize the performance of the solar cells. However, the solar spectrum irradiance and the incident angle received by the solar cells change with the latitude and time. The mismatch between the actual and theoretical solar spectrum could decrease the efficiency of the solar cells. SPCTRL2, a program from National Renewable Energy Laboratory, can provide the detail solar spectrum statistic data in the world which considers the longitude, latitude, atmospheric conditions, and incident angle with a date and a time [15]. All the data of the irradiance of solar spectrum $E(\lambda, \theta)$ and the incident photons flux density $Num(\lambda, \theta)$ are from SPCTRL2 in this paper. With the revolution of the Earth, the incident angle in the south-north direction varies seasonally with the oscillation centre on the day of the spring or autumnal equinox, when the sun is normally incident to the Equator. Hence, the anti-reflective coating optimization which should consider the local solar spectrum of a whole year originally now is simplified to consider the solar spectrum on the day of spring equinox only.

**Table 1. Input parameters used in SPCTRL2 program.**

| | Latitude (°) | Longitude (°) | Aerosol optical depth | Alpha | Albedo (surface reflectance) | Total column ozone (cm) | Total precipitable water vapor (cm) | Surface pressure (mB) | Day of the year |
|---|---|---|---|---|---|---|---|---|---|
| Quito | 0 | 78.5W | 0.27 | 1.14 | 0.2 | 0.34 | 1.42 | 1013.25 | 79 |

| Beijing | 39.9N | 116. 3E | 0.77 | 1.14 | 0.16 | 0.34 | 0.95 | 1040 | 79 |
| Moscow | 55.3N | 37.5E | 0.35 | 1.14 | 0.1 | 0.36 | 1.36 | 750 | 79 |

Figure 2 shows the spectrum of the incident photons flux density of the daytime on the day of the spring equinox for the cities of (a) Quito, (b) Beijing and (c) Moscow according to Eq. (2) when the solar cells are placed on the earth in the horizontal mode. The input parameters in the SPCTRL2 [15] for the cities of Quito, Beijing and Moscow are shown in Table 1, which are all from AERONET [49]. For Quito, the incident angle changes from 90° to 0° then to 90° which corresponds to the sunrise to the noon then to the sunset, because the two-dimensional incident plane remains the same from the sunrise to the sunset. The minimum incident angle is 0°, which is shown in Fig. 2(a). However, for the cities in the southern or northern hemisphere, the two-dimensional incident plane changes all the time from the sunrise to the sunset. The minimal incident angle is determined by the relative position of the Sun and the Earth. The incident angle changes from 90° to 39.9° and then to 90° for Beijing from the sunrise to the sunset, as shown in Fig. 2(b). For the same reason, the minimum incident angle is 55.3° for Moscow, as shown in Fig. 2(c). Therefore, the boundaries of the incident angle in Eq. (1) will change with the actual location of the solar cells. Fig. 2(d) shows the comparison of the actual solar spectrum the solar cell will received over the day of the spring equinox when placed in Quito, Beijing and Moscow, respectively, after integrating over the entire incident angle. The incident photons flux density decreases obviously with the latitude due to the relative position of the sun to the city.

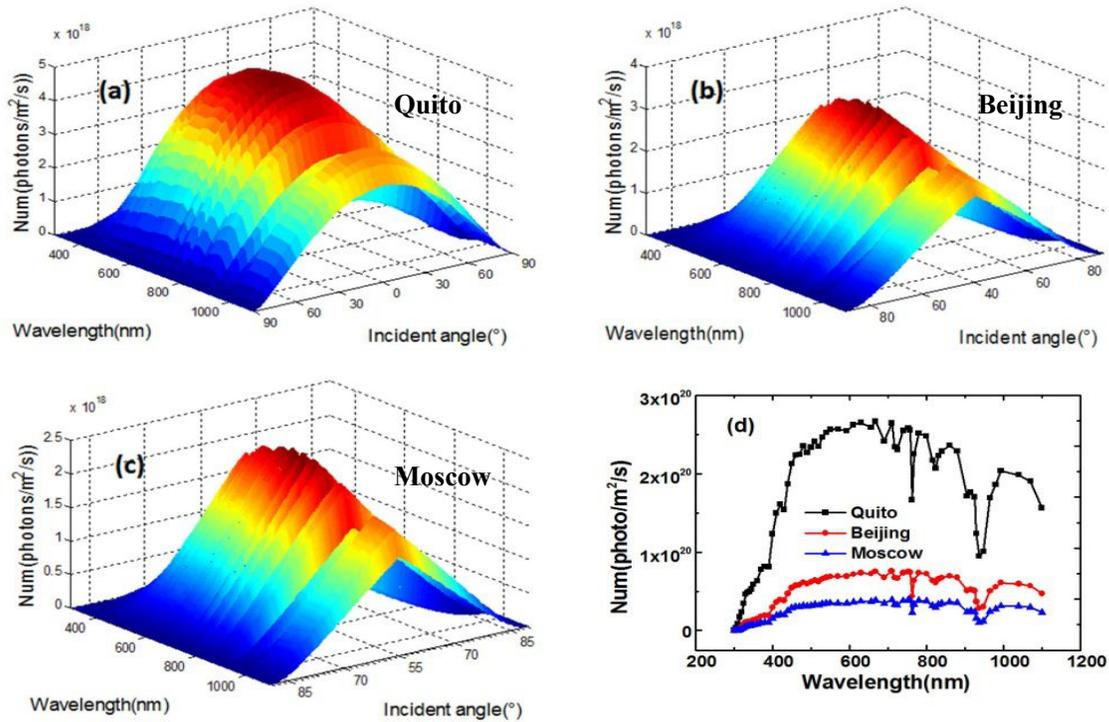

Fig. 2 The dependence of the incident photons flux density on the wavelength and the incident angle on the day of spring equinox for the (a)Quito, (b) Beijing and (c) Moscow, respectively. At noon of the day of spring equinox, the sunlight is vertically incident on the equator. The Quito is located on the equator with 0 degree incident angle at noon of the spring equinox, the Beijing is located 40 degrees north latitude with 40 degree at noon of the day, and Moscow is located 55 degrees north latitude with minimum incident angle of 55 degree. At the sunrise and sunset, the incident angles of these three cities are same 90 degree, maximum incident angle. (d) the comparison of the actual spectrum of the incident photons flux density of three cities at noon of the day of the spring equinox. Based on the atmosphere scatter and absorption of the solar spectrum at different location with its incident angle and latitude, the incident photons flux density have great different on the earth. Therefore, the anti-reflection coating system of the solar cell should be designed with actual flux density for the maximum incident quantum efficiency.

In order to improve the solution accuracy of the anti-reflective coating system, ant colony algorithm with a modified multi-city layer travelling salesman problem model was applied in this work [14,23]. Ant colony algorithm is used to is a heuristic optimization algorithm that mimicking the way of the ants in nature to establish the shortest path from their nest to the food source and then back by leaving their pheromone on the path as a communication medium among them[50,51]. It can enhance the convergence speed and improve possibility to arrive at the optimal solutions [52,53,,54,55,56].. The details of the optimization process for anti-reflective coating by ant colony algorithm were presented in our previous work [14].

**Optimization anti-reflection coating structures:** For the practical materials, the refractive index and the extinction coefficient vary with wavelength, which is known as the dispersion relationship. The calculations presented here, however, only consider single values of refractive index (i.e. not wavelength dependence) and the layers are assumed to be non-absorbing. The actual solar irradiance decreases with the latitude, as illustrated in Fig. 2(d), which indicates the available incident photon number is relatively small. When considering the solar spectrum in different cities and setting $\eta_{in}$ as the evaluation function in the anti-reflective coating optimization, the peak of the quantum efficiency spectrum moves towards around 700nm, which is the peak of the actual solar spectrum and fits the actual solar spectrum very well for all three cities. Table 2 presents the detail structures of two films anti-reflection coating optimized by ant colony algorithm with and without SPCTRL2 incorporated. Figure 3(a)-(d) shows the optical transmittance spectrum with $\lambda$= [300, 1100] nm and $\theta$= [0°, 90°] for the optimization structures of Table 2, wherein $\eta_{in\text{-}Q}$, $\eta_{in\text{-}B}$ and $\eta_{in\text{-}M}$ represent the incident quantum efficiency of Quito, Beijing and Moscow, respectively. First, for comparison, the structure optimized by ant colony algorithm without considering SPCTRL2 is also presented in Figure 3(a), where the area of the contour plot with the transmittance above 80% and 98% are 76.12% and 27.51%, respectively, which are marked by white lines. The $\eta_{in}$ are 95.74%, 92.27%, and 86.78% for Quito, Beijing and Moscow, respectively. Second, the area with higher transmittance redshifts and moves up to larger incident angle in order to adapt to the effect of the location and time in the optimization, which can be distinguished from the contour lines of the transmittance. Compared with Fig. 3(a), the profile of the area with high transmittance changes a lot for the anti-reflective coating optimized with SPCTRL2 incorporated within the same range, as shown in Fig. 3(b)-3(d). The area of the contour plot with the transmittance above 80% and 98% are 73.72% and 35.99%, 71.32% and 27.12%, 69.92% and 17.43%, respectively. However, the $\eta_{in}$ for Quito, Beijing and Moscow are increased to 96.00%, 93.64%, and 91.02%, respectively. In other words, the incident quantum efficiency at Quito, Beijing and Moscow optimized by ant colony algorithm with SPCTRL2 incorporated are 0.26%, 1.37% and 4.24% larger than those optimized by ant colony algorithm without SPCTRL2 incorporated for two-layer anti-reflective coating, respectively. Figure 3(e)-3(g) show the omnidirectional incident quantum efficiencies $\eta_{in}$ ($\lambda$) of Quito, Beijing and Moscow with and without SPCTRL2 incorporated, respectively. It can be seen that, with SPCTRL2 incorporated, the quantum efficiency spectrum almost fits the actual solar spectrum very well at each city. While for the quantum efficiency spectrum optimized without SPCTRL2 incorporated, there is an obvious mismatch between the actual solar spectrum and quantum efficiency spectrum. For example, the transmittance is relatively high at the wavelength of around 400 nm, as shown in the Fig. 3(a). As a result, all the peaks of the quantum efficiency locate at around 400nm for three cities if the solar spectrum was not considered in the optimization, as illustrated in Fig. 3(e)-3(g).

Table 2. Detail structures of two-layer anti-reflective coating optimized by ant colony algorithm with and without SPCTRL2 incorporated.

| | | 1st layer | | 2nd layer | |
|---|---|---|---|---|---|
| optimization anti-reflection coating structures with 2 layers | optimization methods | *refractive index* | *thickness (nm)* | *refractive index* | *thickness (nm)* |
| | No SPCTRL2 | 1.41 | 112.09 | 2.41 | 58.89 |
| | SPCTRL2 at Quito | 1.44 | 113.66 | 2.55 | 60.46 |
| | SPCTRL2 at Beijing | 1.27 | 165.29 | 2.29 | 76.10 |
| | SPCTRL2 at Moscow | 1.17 | 221.62 | 2.15 | 80.80 |

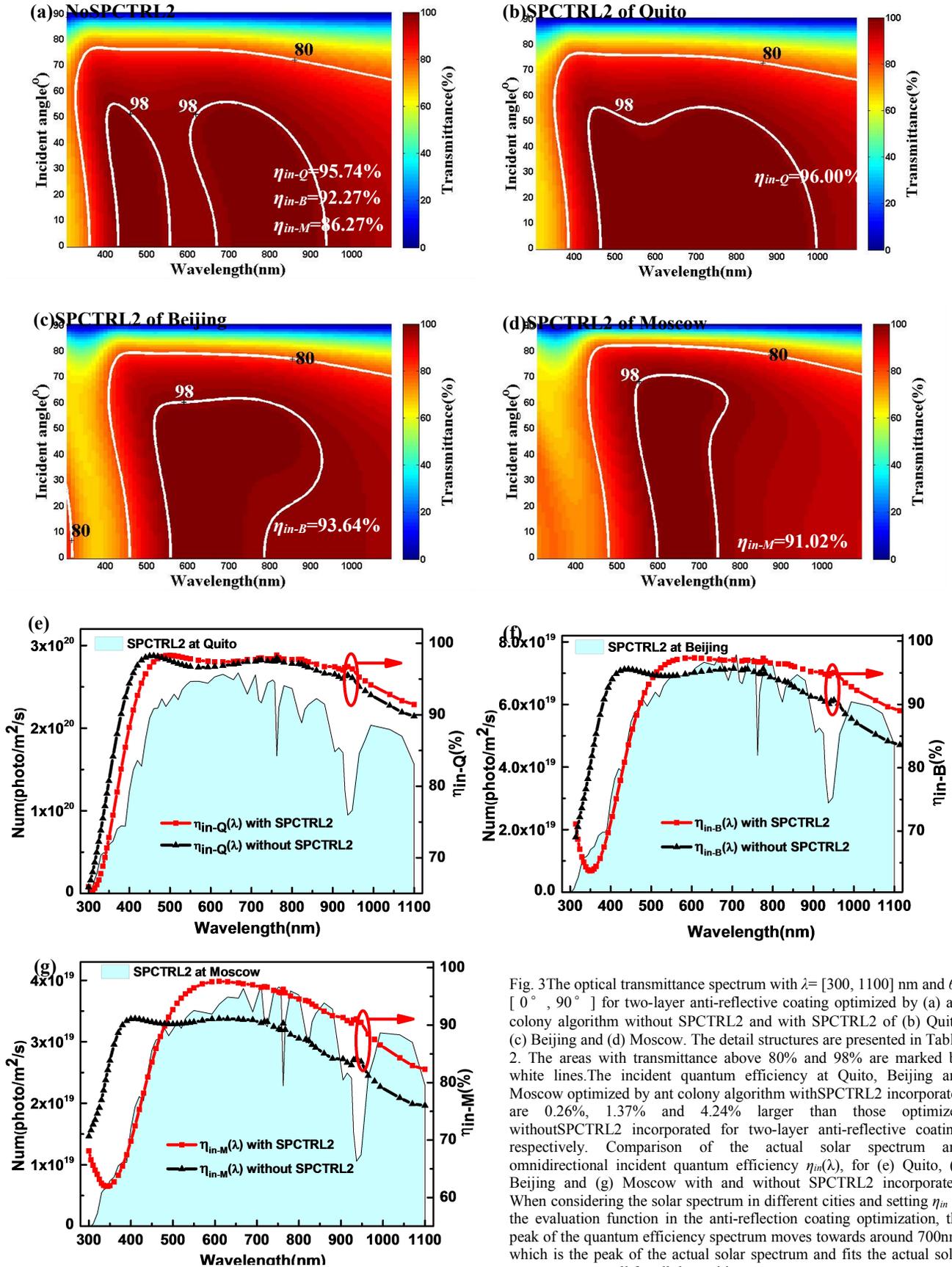

Fig. 3 The optical transmittance spectrum with $\lambda$= [300, 1100] nm and $\theta$= [ 0°, 90° ] for two-layer anti-reflective coating optimized by (a) ant colony algorithm without SPCTRL2 and with SPCTRL2 of (b) Quito, (c) Beijing and (d) Moscow. The detail structures are presented in Table. 2. The areas with transmittance above 80% and 98% are marked by white lines. The incident quantum efficiency at Quito, Beijing and Moscow optimized by ant colony algorithm with SPCTRL2 incorporated are 0.26%, 1.37% and 4.24% larger than those optimized without SPCTRL2 incorporated for two-layer anti-reflective coating, respectively. Comparison of the actual solar spectrum and omnidirectional incident quantum efficiency $\eta_{in}(\lambda)$, for (e) Quito, (f) Beijing and (g) Moscow with and without SPCTRL2 incorporated. When considering the solar spectrum in different cities and setting $\eta_{in}$ as the evaluation function in the anti-reflection coating optimization, the peak of the quantum efficiency spectrum moves towards around 700nm, which is the peak of the actual solar spectrum and fits the actual solar spectrum very well for all three cities.

The detail structures of three-layer anti-reflective coating optimized by ant colony algorithm with and without SPCTRL2 incorporated for Quito, Beijing and Moscow are shown in Table 3. Figure 4(a)-4(d) show the optical transmittance spectra with $\lambda$= [300, 1100] nm and $\theta$= [0°, 90°] for these three-layer anti-reflective coatings. The area of the contour plot with the transmittance above 80% and 98% are 85.20% and 37.66%, 81.14% and 51.92%, 81.37% and 49.12%, 80.64% and 44.78%, which are marked by white lines in Fig. 4(a)-4(d), respectively. Compared with the transmittance spectrum of two-layer anti-reflective coating without SPCTRL2 incorporated, the incident angle with the transmittance larger than 80% expands to 85°. The corresponding $\eta_{in}$ are 96.86%, 95.72%, and 94.46% for Quito, Beijing and Moscow, respectively. However, the actual $\eta_{in}$ could increase to 97.60%, 97.03%, and 95.74%, respectively, if the anti-reflective coating optimization with SPCTRL2 incorporated. That is to say, the incident quantum efficiency at Quito, Beijing and Moscow optimized by ant colony algorithm with SPCTRL2 are 0.74%, 1.31% and 1.28% higher than that optimized without SPCTRL2 for three-layer anti-reflective coating, respectively. Fig. 4(e)-4(g) compare the actual solar spectrum and the dependence of the $\eta_{in}$ on the wavelength for Quito, Beijing and Moscow with and without SPCTRL2 incorporated, respectively. Both of the quantum efficiency spectra of three-layer anti-reflective coating optimized with and without SPCTRL2 incorporated improve a lot. At the short wavelength region, both of the $\eta_{in}$ are almost the same. However, for the wavelength longer than 800nm, the $\eta_{in}$ considered SPCTRL2 is a little bit larger than that without considering SPCTRL2. Moreover, compared with the incident quantum efficiency considering SPCTRL2, it can be concluded that it is important for the high latitude cities to incorporate the location-specific solar spectrum during the design of the solar cells.

**Table 3. Detail structures of three-layer anti-reflection coating optimized by ant colony algorithm with and without the SPCTRL2 incorporated.**

| | | 1st layer | | 2nd layer | | 3nd layer | |
|---|---|---|---|---|---|---|---|
| optimization anti-reflection coating structures with 3 layers | optimization methods | *refractive index* | *thickness (nm)* | *refractive index* | *thickness (nm)* | *refractive index* | *thickness (nm)* |
| | No SPCTRL2 | 1.06 | 370.27 | 1.44 | 113.66 | 2.55 | 57.33 |
| | SPCTRL2 at Quito | 1.15 | 159.04 | 1.60 | 102.71 | 2.65 | 57.33 |
| | SPCTRL2 at Beijing | 1.09 | 248.22 | 1.55 | 110.53 | 2.61 | 60.46 |
| | SPCTRL2 at Moscow | 1.08 | 274.82 | 1.45 | 124.61 | 2.51 | 65.15 |

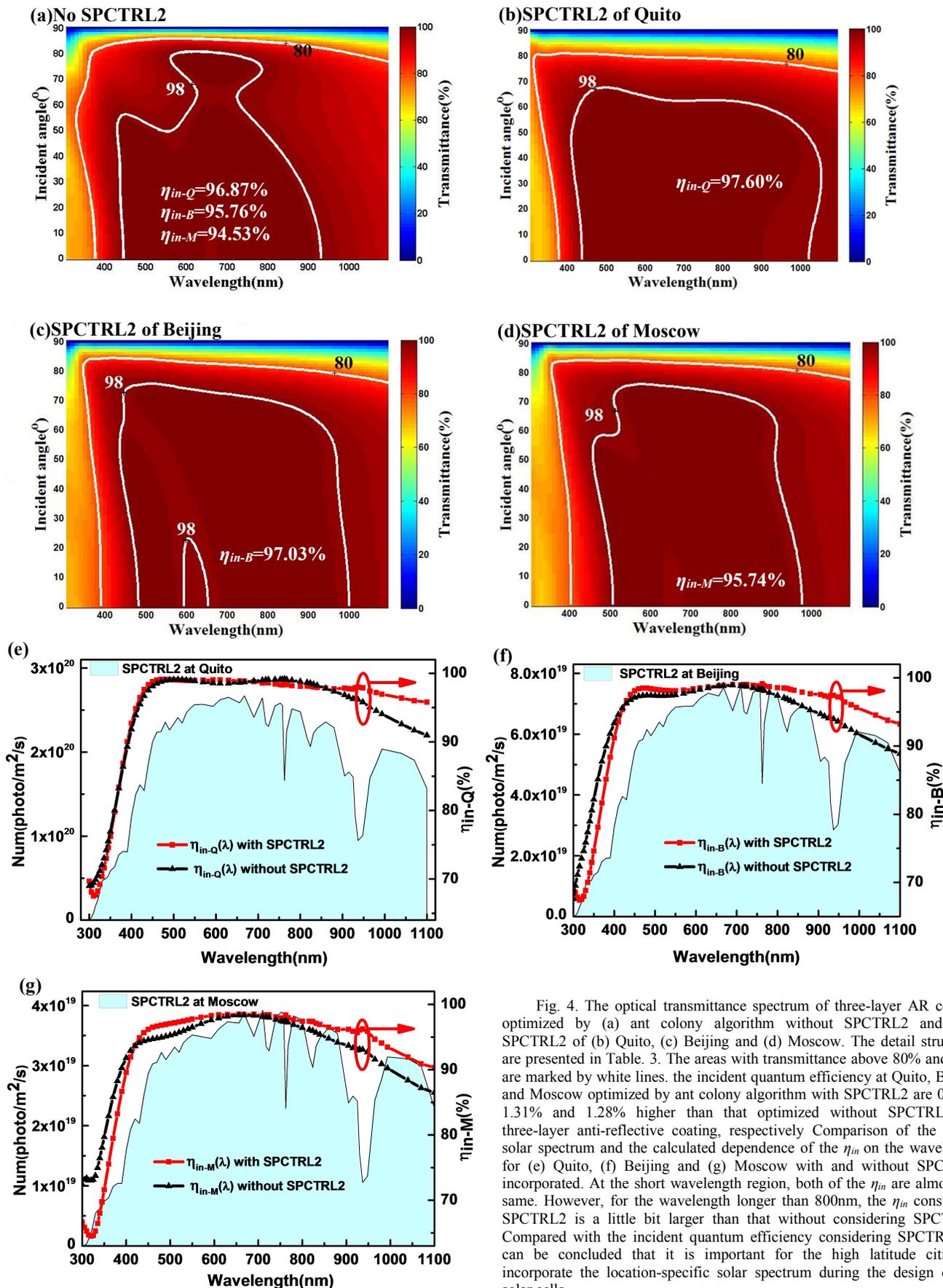

Fig. 4. The optical transmittance spectrum of three-layer AR coating optimized by (a) ant colony algorithm without SPCTRL2 and with SPCTRL2 of (b) Quito, (c) Beijing and (d) Moscow. The detail structures are presented in Table. 3. The areas with transmittance above 80% and 98% are marked by white lines. the incident quantum efficiency at Quito, Beijing and Moscow optimized by ant colony algorithm with SPCTRL2 are 0.74%, 1.31% and 1.28% higher than that optimized without SPCTRL2 for three-layer anti-reflective coating, respectively Comparison of the actual solar spectrum and the calculated dependence of the $\eta_{in}$ on the wavelength for (e) Quito, (f) Beijing and (g) Moscow with and without SPCTRL2 incorporated. At the short wavelength region, both of the $\eta_{in}$ are almost the same. However, for the wavelength longer than 800nm, the $\eta_{in}$ considered SPCTRL2 is a little bit larger than that without considering SPCTRL2. Compared with the incident quantum efficiency considering SPCTRL2, it can be concluded that it is important for the high latitude cities to incorporate the location-specific solar spectrum during the design of the solar cells.

In practice, only refractive index values of bulk materials between 1.46- 2.55 are available for common silicon solar cell, which limited the current commercial availability and use of low-refractive index materials in solar cell applications. Two-layer and three-layer anti-reflective coatings were optimization with restriction of refractive index values between 1.46- 2.55 by ant colony algorithm with and without SPCTRL2 incorporated, and the detail structures and compares $\eta_{in}$ of two-layer and three-layeranti-reflective coating are shown in Table 4 and Table 5, respectively. However, compared the refractive index ranging from 1.05 to 2.66 and from 1.46 to 2.55, the $\eta_{in}$ at Quito, Beijing and Moscow optimized by ant colony algorithm without and with SPCTRL2 incorporated are 96.00% and 95.30%, 93.64% and 92.83%, 91.02% and 87.71%, respectively. And it is 97.60% and 95.96%, 97.03% and 93.02%, 95.74% and 87.94% for three-layer anti-reflective coating, respectively. The effect is still obvious. It is necessary to do the careful and nuanced optimizations for anti-reflective coatings even in practice with restrictions on the refractive index selection, especially for the cities in the high latitude. Comparing the detail structure parameters, it was found that the lower limit of the refractive index is more important for the $\eta_{in}$, because the difference of the refractive index between the air and solar cell interface determines the reflection. Expanding the range of the refractive index for the practical material can be achieved by nano-technology and advanced deposition technique[6,12], which is proved to be an useful technique to manipulate the light coupling in silicon photonics [29,30].

Table 4. Detail structures of two-layer anti-reflective, which composes of bulk materials with refractive index values between 1.46- 2.55, coating optimized by ant colony algorithm with and without SPCTRL2 incorporated. The corresponding $\eta_{in-Q}$, $\eta_{in-B}$ and $\eta_{in-M}$ over $\lambda$= [300, 1100]nm and $\theta$= [0°,90°] are also shown.

| | optimization methods | 1st layer | | 2nd layer | | $\eta_{in-Q}$ | $\eta_{in-B}$ | $\eta_{in-M}$ |
| --- | --- | --- | --- | --- | --- | --- | --- | --- |
| | | refractive index | thickness (nm) | refractive index | thickness (nm) | | | |
| optimization anti-reflection coating structures with 2 layers | No SPCTRL2 | 1.46 | 90.19 | 2.45 | 49.51 | 94.37% | 89.76% | 83.56% |
| | SPCTRL2 at Quito | 1.47 | 99.57 | 2.49 | 68.58 | 95.30% | - | - |
| | SPCTRL2 at Beijing | 1.48 | 121.48 | 2.51 | 65.15 | - | 92.83% | - |
| | SPCTRL2 at Moscow | 1.46 | 151.21 | 2.54 | 65.15 | - | - | 87.71% |

Table 5.Detail structures of three-layer anti-reflective layers, which compose of bulk materials with refractive index values between 1.46 and 2.55, coating optimized by ant colony algorithm with and without SPCTRL2 incorporated. The corresponding $\eta_{in-Q}$, $\eta_{in-B}$ and $\eta_{in-M}$ over $\lambda$= [300, 1100]nm and $\theta$= [0°,90°] are also shown.

| optimization anti-reflection coating structures with 3layers | optimization methods | 1st layer | | 2nd layer | | 3nd layer | | $\eta_{in\text{-}Q}$ | $\eta_{in\text{-}B}$ | $\eta_{in\text{-}M}$ |
|---|---|---|---|---|---|---|---|---|---|---|
| | | *refractive index* | *thickness (nm)* | *refractive index* | *thickness (nm)* | *refractive index* | *thickness (nm)* | | | |
| | No SPCTRL2 | 1.46 | 83.92 | 2.03 | 58.89 | 2.55 | 51.07 | 94.47% | 90.91% | 85.26% |
| | SPCTRL2 at Quito | 1.46 | 102.70 | 2.17 | 11.95 | 2.50 | 52.63 | 95.96% | - | - |
| | SPCTRL2 at Beijing | 1.46 | 82.36 | 1.47 | 51.07 | 2.53 | 65.15 | - | 93.02% | - |
| | SPCTRL2 at Moscow | 1.46 | 126.17 | 2.17 | 19.77 | 2.49 | 54.20 | - | - | 87.94% |

**Discussion**

In this paper, the incident quantum efficiency $\eta_{in}$ is set as evaluating function with SPCTRL2 program incorporated to optimize the broadband omnidirectional anti-reflective coating by ant colony algorithm method for silicon solar cells for the purpose of the practical applications. The solar spectrum on the day of the spring equinox was selected in order to expand the one-day optimization to a whole year optimization. Two-layer and three-layer anti-reflective coating are optimized over $\lambda$ = [300, 1100] nm and $\theta$ = [0°, 90°]. The $\eta_{in}$ of two-layer anti-reflective coating optimized by ant colony algorithm with SPCTRL2 incorporated increase by 0.26%, 1.37% and 4.24% larger than that without SPCTRL2 incorporated in Quito, Beijing and Moscow, respectively. While for three-layer anti-reflective coating, the $\eta_{in}$ are improved by 0.74%, 1.31% and 1.28%, respectively. The anti-reflective coating optimized by ant colony algorithm method with SPCTRL2 incorporated can further improve the efficiency of the solar cells, especially for the high latitude locations.

In practice, surface texture or even encapsulation may be applied to the real solar cells. The refractive index of the incident material may be changed. Also, the encapsulation which comprising a glass cover will change the angle range of light incident on the cell over a day due to the refraction at the air/glass interface. For the practical purpose, this optimization method can be utilized by modifying a little bit in the form of optimization boundary conditions, due to the changing of refractive index and thickness distribution after surface texture and encapsulation. This work paves a way for solar cell with high quantum efficiency, leading to a high utilization efficiency and high light absorption of solar irradiance at different location of the global by the optimization of anti-reflection coating, and thus demonstrates significant progress in "green energy with high calorific value".

**Acknowledgements**

This work was supported by the National Natural Science Foundation of China (Grant No. 61222501, 61335004 and 61505003), the china Postdoctoral Science Foundation funded project (Project No.: 143930), the Specialized Research Fund for the Doctoral Program of Higher Education of China (Grant No. 20111103110019) and the Postdoctoral Science Foundation of Beijing funded project (Grant No. Q6002012201502).

**Author contributions**

X.G., Q. L. and C.L. conceived the study. H. Z. and B. L. conducted the mathematical calculation. In the meanwhile, Y F and H. W. carried out the simulation. X.G. and C.X. supervised the entire research project. X.G., Q. L., C.L., H. Z., B. L. Y. F H. W and W. L. analysed the data. All authors discussed the results and wrote the manuscript.

**Additional information**

Competing financial interests: The authors declare no competing financial interests.